\documentclass[aps,prm,twocolumn,superscriptaddress]{revtex4-2}
\usepackage{gensymb}
\usepackage{graphicx}
\usepackage{booktabs}
\usepackage{multirow}
\usepackage{chemformula}
\newcommand{\lntao}{\mbox{\textit{Ln}TaO$_4$}}
\newcommand{\lnnbo}{\mbox{\textit{Ln}NbO$_4$}}
\begin{document}
\title{Magnetism on the stretched diamond lattice in lanthanide orthotantalates}
\author{Nicola D.~Kelly}
\email{ne281@cam.ac.uk}
\author{Lei Yuan}
\author{Rosalyn L.~Pearson}
\affiliation{Cavendish Laboratory, University of Cambridge, J J Thomson Avenue, Cambridge, CB3 0HE, UK}
\author{Emmanuelle Suard}
\affiliation{Institut Laue-Langevin, 71 Avenue des Martyrs, 38000 Grenoble, France}
\author{In\'{e}s Puente~Orench}
\affiliation{Institut Laue-Langevin, 71 Avenue des Martyrs, 38000 Grenoble, France}
\affiliation{Instituto de Nanociencia y Materiales de Arag\'{o}n (INMA), CSIC-Universidad de Zaragoza, Zaragoza, 50009, Spain}
\author{Si\^{a}n E.~Dutton}
\email{sed33@cam.ac.uk}
\affiliation{Cavendish Laboratory, University of Cambridge, J J Thomson Avenue, Cambridge, CB3 0HE, UK}
\date{\today}
\begin{abstract}
The magnetic \textit{Ln}$^{3+}$ ions in the fergusonite and scheelite crystal structures form a distorted or stretched diamond lattice which is predicted to host exotic magnetic ground states. In this study, polycrystalline samples of the fergusonite orthotantalates $M$-\lntao\ (\textit{Ln}~= Nd, Sm, Eu, Gd, Tb, Dy, Ho, Er) are synthesized and then characterized using powder diffraction and bulk magnetometry and heat capacity. \ch{TbTaO4} orders antiferromagnetically at 2.25~K into a commensurate magnetic cell with $\vec{k}=0$, magnetic space group 14.77 ($P2_1$$'/c$) and Tb moments parallel to the \textit{a}-axis. No magnetic order was observed in the other materials studied, leaving open the possibility of exotic magnetic states at $T<2$~K.
\end{abstract}
\maketitle
\section{Introduction}
Magnetism on diamond-like lattices has been widely studied in both coordination frameworks \cite{Abdeldaim2020} and ceramic systems, including materials with the scheelite crystal structure such as \ch{KRuO4} \cite{Marjerrison2016} and \ch{KOsO4} \cite{Injac2019} as well as cubic spinels $AB_2$O$_4$ with a magnetic ion on the $A$-site \cite{Tristan2005,Chen2009,Zaharko2011,Ge2017}. The perfect diamond lattice is bipartite and unfrustrated, expected to order into a collinear antiferromagnetic ground state if only nearest-neighbor interactions ($J_1$) are considered \cite{MacDougall2011}. However, magnetic frustration can arise if interactions with the twelve next-nearest-neighbors ($J_2$) are included, or if distortion lowers the symmetry from cubic. This may give rise to exotic magnetic behavior including spiral spin-liquid states \cite{Bergman2007,Gao2017} or topological paramagnetism \cite{Wang2015,Chen2017,Chamorro2018}.

\begin{figure}[htbp]
\centering
\includegraphics[width=8.6cm]{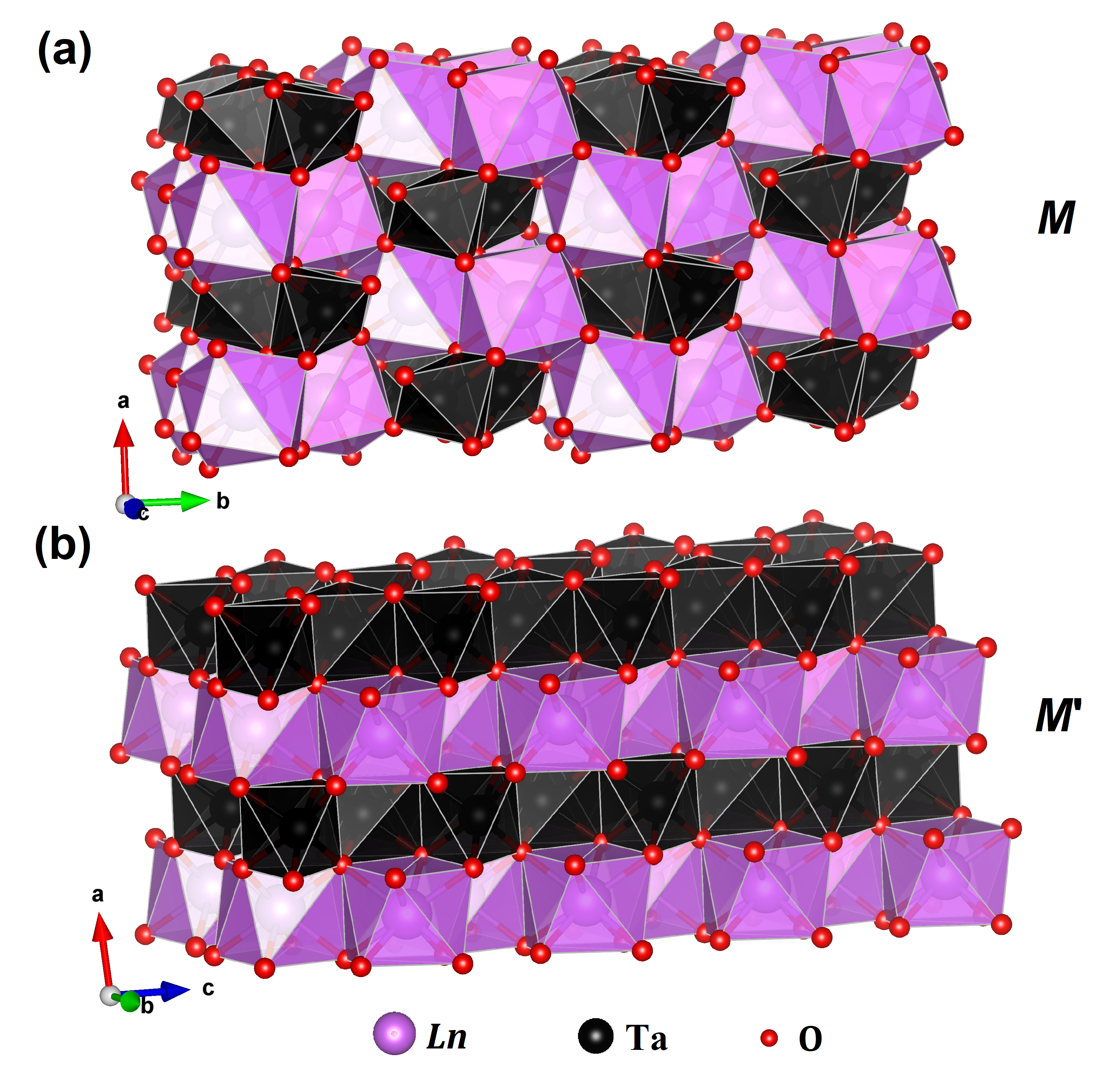}
\caption{Crystal structures of (a) $M$ ($I2/a$), (b) $M'$ ($P2/c$) phases of \lntao.}
\label{fig:lntaostructures}
\end{figure}

Rare-earth orthoniobates \lnnbo\ and orthotantalates \lntao\ (\textit{Ln}~= Y, La--Lu) are of wide interest as a result of their luminescent \cite{Hristea2009,Voloshyna2013}, proton-conducting \cite{Haugsrud2006}, oxide-ion-conducting \cite{Li2014} and dielectric properties \cite{Kim2006}. The tantalates have also been proposed as thermal barrier coatings for gas turbines \cite{Wang2017}. The niobates and tantalates share two common crystallographic polymorphs: fergusonite ($I2/a$, monoclinic, $M$) and scheelite ($I4_1/a$, tetragonal, $T$) \cite{Komkov1959,Keller1962,Rooksby1963,Wolten1967,Brixner1983,Mather1995,Ryumin2017,Saura-Muzquiz2021}. Additionally, the tantalates may crystallize in different monoclinic and tetragonal ($M'$, $T'$) phases depending on the synthesis conditions. The $T$ phase has been observed using \textit{in situ} diffraction experiments, but it rapidly converts to the $M$ phase upon cooling and cannot be isolated at room temperature. The $M$-$T$ transformation temperature occurs at 1325--1410~\degree C for the tantalates and 500--800~\degree C for the niobates; within each series this transition temperature increases with decreasing \textit{Ln}$^{3+}$ radius \cite{Stubican1964}. The two monoclinic polymorphs of \lntao\ are closely related: to change from $M$ to $M'$ only involves halving the $b$-axis and removing the body-centering \cite{Wolten1967,Hartenbach2005}. The metal--oxygen coordination polyhedra (distorted square antiprisms for \textit{Ln}$^{3+}$ and distorted octahedra for Ta$^{5+}$) are approximately the same in both phases \cite{Brixner1983}. However, the change of centering means that the $M$ and $M'$ structures have significantly different arrangements of the polyhedral building blocks: distinct layers perpendicular to $a$ in the $M'$ phase, but a different, non-layered arrangement in the $M$ phase \cite{Mather1995} as shown in Fig.~\ref{fig:lntaostructures}. The arrangement of lanthanide ions in both the $M$ and $T$ phases is equivalent to the arrangement of carbon atoms in diamond, but distorted or `stretched' \cite{Bordelon2021} into lower symmetry (monoclinic or tetragonal).

Numerous diffraction studies have been carried out on the lanthanide niobates and tantalates with both powder and single-crystal samples \cite{Santoro1980,Tsunekawa1993a,Hartenbach2005,Sarin2014,Saura-Muzquiz2021} but their magnetic properties are under-researched by comparison. In 1965, Wang and Gravel measured the susceptibilities of $M$-\ch{GdNbO4} and $M$-\ch{NdNbO4} at 5--750~K and found paramagnetism and antiferromagnetism respectively, with $T_\mathrm{N}=25$~K for \ch{NdNbO4} \cite{Wang1965}. Cashion \textit{et al}.~investigated $M$-\lnnbo\ with \textit{Ln}~= Nd, Gd, Tb, Dy, Ho, Er and Yb but found magnetic transitions only below 2~K (i.e.~no feature at 25~K in \ch{NdNbO4}), again with negative Curie-Weiss temperatures indicating antiferromagnetic interactions \cite{Cashion1968}. At around the same time Starovoitov \textit{et al}.~independently measured isothermal magnetization on $M$-\lnnbo\ with \textit{Ln}~= Nd, Eu, Sm, Gd, Dy, Ho and Yb, finding evidence for single-ion anisotropy in all samples except \ch{GdNbO4}, as expected for a $f^7$ system with $L=0$ \cite{Starovoitov1969}. Tsunekawa \textit{et al}.~measured the susceptibility of single crystals of selected $M$-\lnnbo\ (\textit{Ln}~= Nd, Gd, Ho) and $M$-\lntao\ (\textit{Ln}~= Nd, Ho, Er) in the range 4.2 to 300~K; again $\theta_\mathrm{CW}<0$ for all compounds, with no magnetic ordering observed. Deviations from the Curie-Weiss law at low temperatures and a marked anisotropy in the susceptibility were observed in all cases except \ch{GdNbO4}; this behavior was attributed to a crystal field with its principal direction along the $c$-axis, with the greatest effect occurring for \textit{Ln}~= Nd \cite{Tsunekawa1993,Tsunekawa1996}.

This article reports the bulk magnetic characterization of eight powder samples in the series $M$-\lntao, \textit{Ln}~= Nd, Sm, Eu, Gd, Tb, Dy, Ho, Er. The compounds with the larger lanthanides \textit{Ln}~= La, Ce and Pr do not form the $M$ structure \cite{Cava1981,Hartenbach2005,Saura-Muzquiz2021} and were therefore excluded from this study. We confirm the absence of long-range ordering in \ch{NdTaO4}, \ch{HoTaO4} and \ch{ErTaO4} above 2~K, extending the range investigated in a previous study \cite{Tsunekawa1996}. \ch{TbTaO4} was also studied using high-resolution powder neutron diffraction. It exhibits a transition at $T=2.25$~K to a commensurate antiferromagnetic structure with $\vec{k}=0$. No magnetic transitions occurred above 2~K for the remaining samples with \textit{Ln}~= Sm, Eu, Gd, Dy. These measurements indicate the presence of magnetic frustration in $M$-\lntao\ and lay the foundations for future investigations, as possible hosts for novel magnetic states as predicted for the stretched diamond magnetic lattice.

\section{Experimental}
Polycrystalline samples of $M$-\lntao\ were synthesized according to a ceramic procedure, starting from \ch{Ta2O5} (Alfa Aesar, 99.993~\%) and \textit{Ln}$_2$O$_3$ (\textit{Ln}~= Nd, Sm, Eu, Gd, Dy, Ho, Er, Y) or \ch{Tb4O7} (Alfa Aesar; all lanthanide oxides $\geq$~99.99~\%). Lanthanide oxides were dried in air at 800~\degree C overnight before weighing. For each compound, 1:1 molar amounts of the reagents were ground with an agate pestle and mortar, pressed into a 13~mm pellet and placed in an alumina crucible. Pellets were fired for 3~x 24~h at 1500~\degree C in air with intermediate regrinding. The exception was \ch{ErTaO4}, which first formed the $M'$ phase ($P2/c$) at 1500~\degree C and required an additional 2~x 24~h at an elevated temperature, 1600~\degree C, to form solely the desired $M$ phase. Heating and cooling rates were 3~\degree C per minute.

Powder X-ray diffraction (PXRD) was carried out at room temperature on a Bruker D8 diffractometer (Cu K$\alpha$, $\lambda = 1.541$ \AA) in the range $10 \leq 2\theta(\degree) \leq 70$ with a step size of 0.02\degree, 0.6 seconds per step. Rietveld refinements \cite{Rietveld1969} were carried out using \textsc{Topas} \cite{Coelho2018} with a Chebyshev polynomial background and a modified Thompson-Cox-Hastings pseudo-Voigt peak shape \cite{Young1993}. \textsc{Vesta} \cite{Momma2011} was used for crystal structure visualization and production of figures.

Powder neutron diffraction (PND) was carried out on a 3~g sample of \ch{TbTaO4} on the D1B and D2B diffractometers (high intensity and high resolution respectively), ILL, Grenoble, using an Orange cryostat ($1.5\leq T(K)\leq300$). Wavelengths were refined to 2.52461(6)~\AA\ for D1B and 1.594882(10)~\AA\ for D2B. Determination of the magnetic structure was carried out using \textsc{FullProf} \cite{Rodriguez-Carvajal1993} and \textsc{Topas} \cite{Coelho2018}. The background was modelled with a Chebyshev polynomial and the peak shape modelled with a modified Thompson-Cox-Hastings pseudo-Voigt function with axial divergence asymmetry \cite{Young1993}.

DC magnetization was measured on warming on a Quantum Design MPMS 3 at a field of 500~Oe in the temperature range $2 \leq T$(K) $\leq 300$, after cooling from 300~K in zero applied field (ZFC) or 500~Oe applied field (FC). Isothermal magnetization was measured on a Quantum Design PPMS DynaCool using the ACMS-II option in its DC magnetometer mode in the field range $\mu_0H$~= 0--9~T. In a low field, up to 500 Oe, the $M(H)$ curve was linear and the susceptibility could therefore be approximated by $\chi(T)=M/H$.

Zero-field heat capacity of \ch{TbTaO4} was measured on the PPMS in the range $1.8\leq T$(K) $\leq 30$. The sample was mixed with an equal mass of Ag powder (Alfa Aesar, 99.99~\%, --635 mesh) to improve thermal conductivity, then pressed into a 1~mm thick pellet for measurement. Apiezon N grease was used to provide thermal contact between the sample platform and the pellet. Fitting of the relaxation curves was done using the two-tau model. The contribution of Ag to the total heat capacity was subtracted using scaled values from the literature \cite{Smith1995}. The \ch{TbTaO4} lattice contribution was estimated and subtracted using a Debye model with $\theta_\mathrm{D}=370$~K \cite{Gopal1966}. 

\section{Results}
\subsection{Crystal structure}
For \textit{Ln}~= Nd--Ho \&\ Y, a small amount, $<5$~wt~\%, of the metastable $M'$-phase (space group $P2/c$; Fig.~\ref{fig:lntaostructures}(b)) was formed in the first heating step but disappeared on further heating. \ch{ErTaO4} formed only the $M'$ phase at 1500~\degree C but formed the desired $M$ phase after heating at 1600~\degree C. Attempts to produce $M$-\ch{YbTaO4} by the same methods were unsuccessful, in agreement with previous authors who found that making this phase requires quenching from high temperature and/or pressure \cite{Markiv2002,Wang2017}. Synthesis of \ch{YbTaO4} by spark plasma sintering (SPS) was attempted as reported in the literature (various experiments with $T\leq 1600$~\degree C, $p\leq 500$ bar, fast or slow cooling \cite{Wu2019}) but was unsuccessful, producing only the $M'$ phase with unreacted \ch{Yb2O3} and \ch{Ta2O5}. It thus appears that the relative stability of $M'$ over $M$ increases across the lanthanide series with decreasing ionic radius, since \ch{LuTaO4} also favors the $M'$ phase \cite{Wang2017} and the solid solution Y$_{1-x}$Yb$_x$TaO$_4$ favors $M'$ when $x\geq0.5$ \cite{Wu2020}.

\begin{figure}[htbp]
\centering
\includegraphics[width=8.6cm]{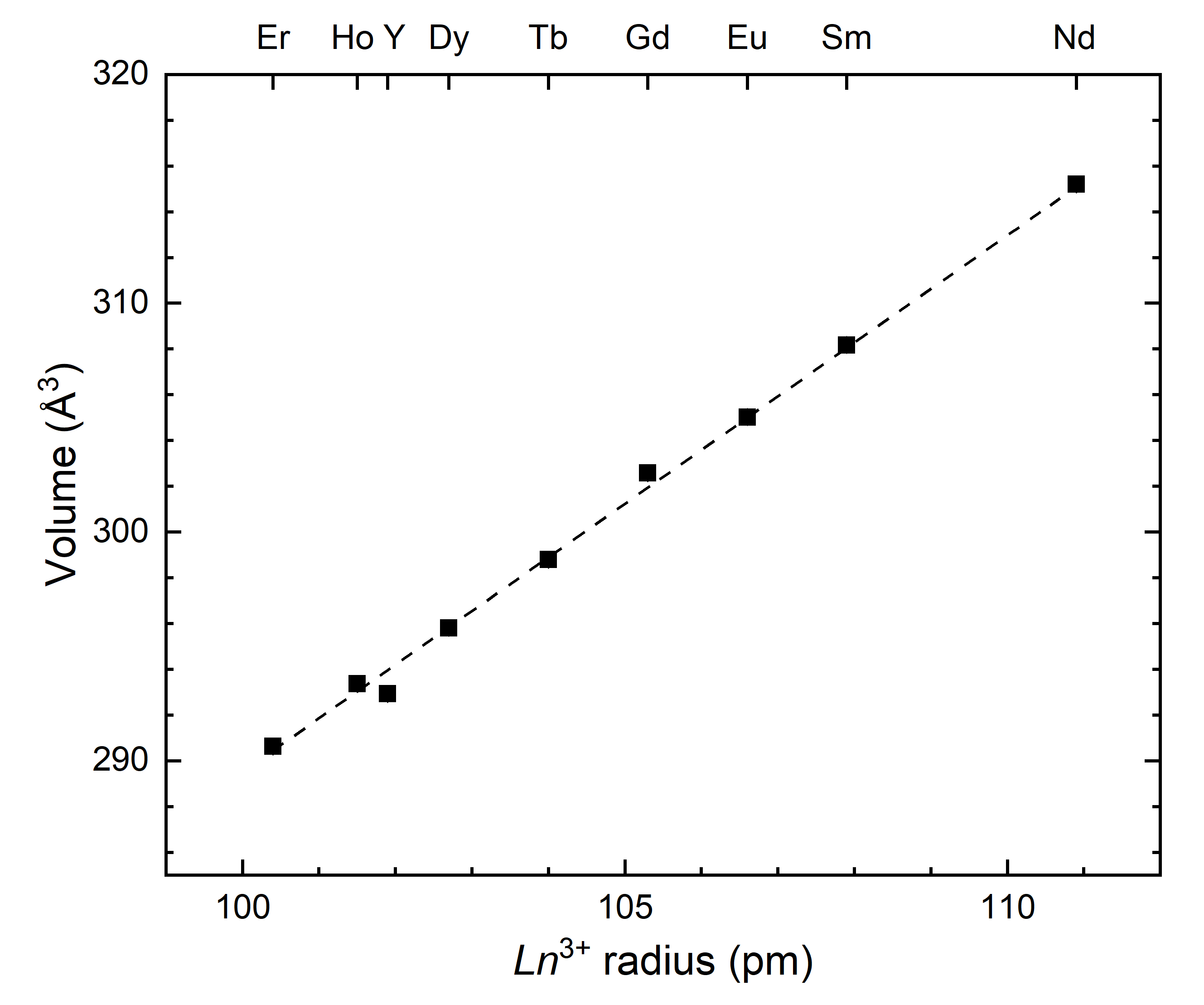}
\caption{Unit cell volumes of $M$-\lntao\ compounds after PXRD and Rietveld refinement with dashed line of best fit to guide the eye. Error bars are smaller than the datapoints. Ionic radius is for an 8-coordinate ion \cite{Shannon1976}.}
\label{fig:lntaovegard}
\end{figure}

PXRD and Rietveld refinement indicated that each sample eventually formed a single phase with the monoclinic $M$-\lntao\ crystal structure, space group $I2/a$, shown in Fig.~\ref{fig:lntaostructures}(a). Unit cell dimensions and the \textit{Ln}$^{3+}$ and Ta$^{5+}$ atomic positions were refined, but the positions of O$^{2-}$ ions were fixed at values taken from neutron diffraction of \ch{NdTaO4} \cite{Santoro1980} because of the low X-ray scattering power of oxygen compared with the heavier metal ions. Refinement of fractional site occupancies with fixed overall stoichiometry indicated that there was no disorder between the \textit{Ln}$^{3+}$ and Ta$^{5+}$ cations. This result is as expected because 6-coordinate Ta$^{5+}$ is much smaller than any of the 8-coordinate lanthanide ions \cite{Shannon1976}. Refined unit cell parameters (Supplemental Material) are in good agreement with literature results \cite{Keller1962,Santoro1980,Wang2017}. A representative Rietveld fit is shown in Fig.~\ref{fig:ndtaofit}; fits for the remaining compounds are in the Supplemental Material. The unit cell volume decreased linearly with decreasing lanthanide ionic radius (Fig.~\ref{fig:lntaovegard}).

\begin{figure}[htbp]
\centering
\includegraphics[width=8.6cm]{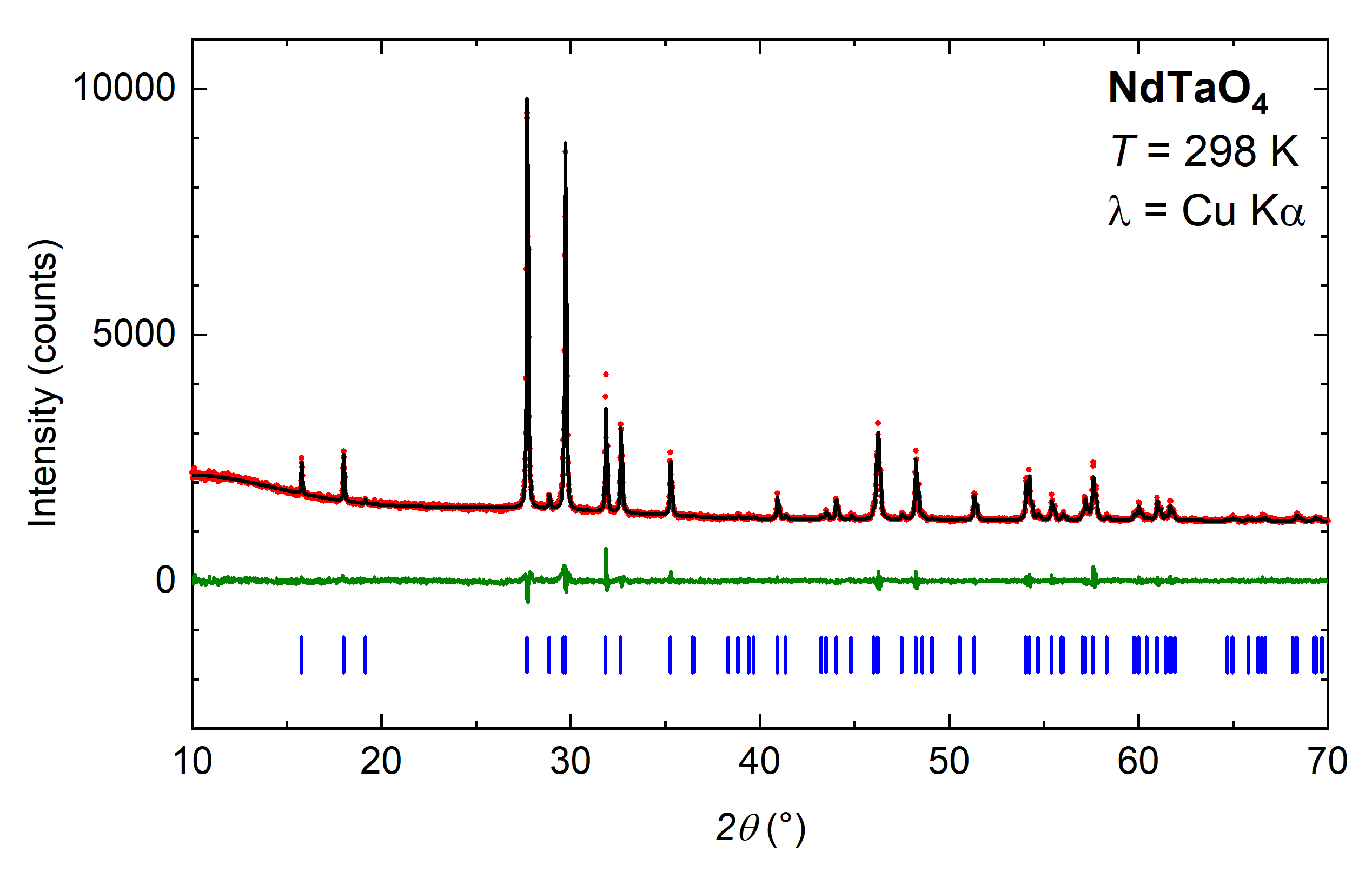}
\caption{Room temperature PXRD pattern for \ch{NdTaO4}: red dots -- experimental data; black line -- calculated intensities; green line -- difference pattern; blue tick marks -- Bragg reflection positions.}
\label{fig:ndtaofit}
\end{figure}

\begin{figure}[htbp]
\centering
\includegraphics[width=8.6cm]{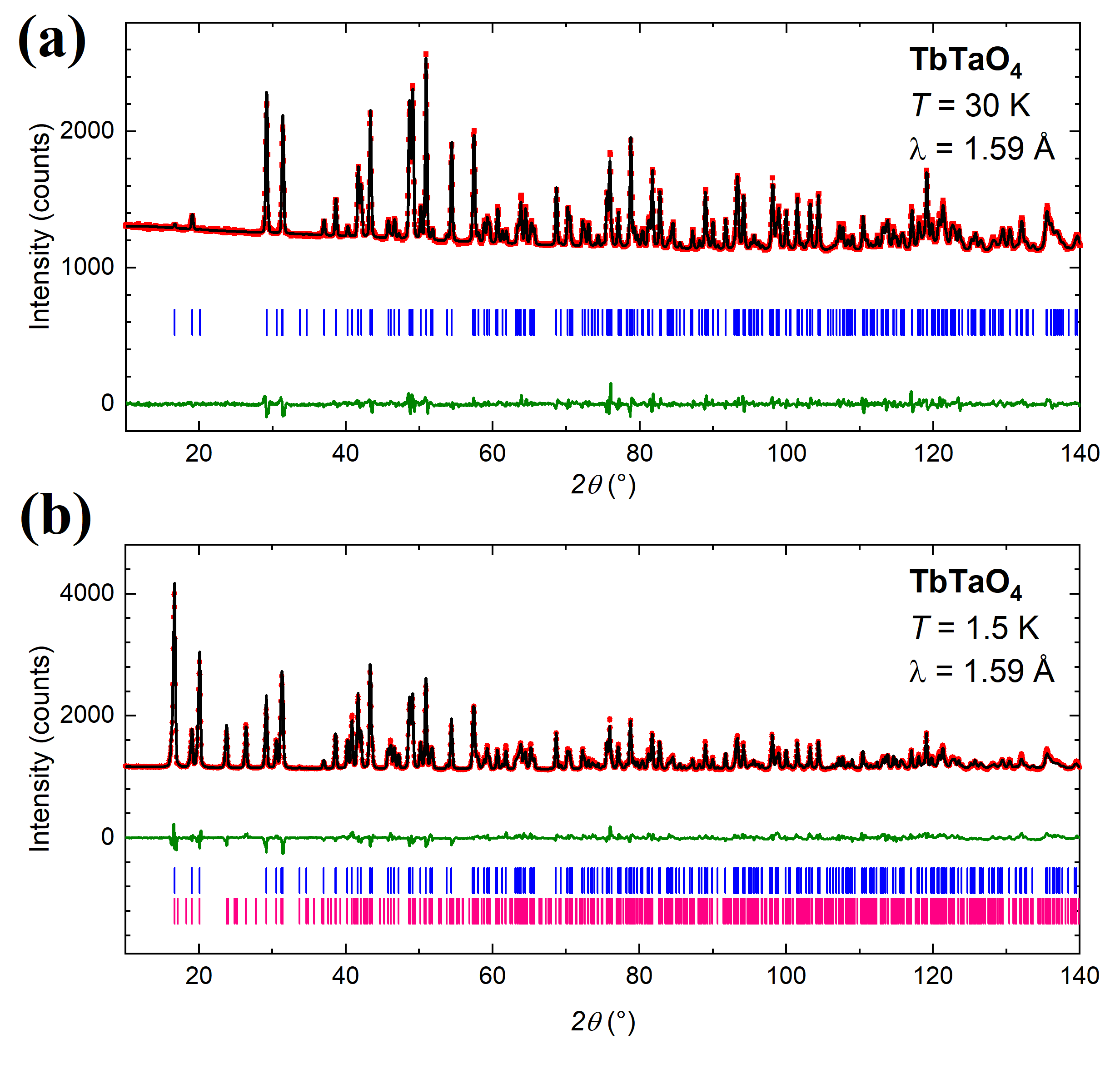}
\caption{PND data for \ch{TbTaO4} collected at $\lambda=1.59$~\AA\ on the D2B diffractometer: (a) 30~K, (b) 1.5~K. Red dots -- experimental data; black line -- calculated intensities; green line -- difference pattern; tick marks -- nuclear (blue) and magnetic (pink) Bragg reflection positions.}
\label{fig:tbtaopnd}
\end{figure}

We obtained high-resolution powder neutron diffraction (PND) data for the paramagnetic phase of \ch{TbTaO4} at 30~K using the D2B beamline at the ILL \cite{Kelly2021ILL}. The measurements confirmed that the nuclear structure of \ch{TbTaO4} is consistent with previous reports for \lntao\ compounds \cite{Keller1962,Santoro1980,Saura-Muzquiz2021}. Fig.~\ref{fig:tbtaopnd}(a) shows a Rietveld refinement of PND data collected at $T=30$~K with $\lambda\approx1.59$~\AA. Interatomic distances were also obtained. The Ta$^{5+}$ ions are surrounded by four shorter and two longer Ta--O bonds, forming octahedra distorted by a second-order Jahn-Teller effect \cite{Saura-Muzquiz2021}, while the Tb$^{3+}$ ions are 8-coordinate. The refined bond lengths are listed in Table~\ref{table:bondlengths}.

\begin{table}[htbp]
\caption{Refined interatomic distances for \ch{TbTaO4} from PND data collected at 1.59~\AA\ (D2B, ILL).}
\label{table:bondlengths}
\begin{ruledtabular}
\begin{tabular}{c c c c}
Atoms & & \multicolumn{2}{c}{Distance /\AA} \\
 & & $T=1.5$~K & $T=30$~K \\
\midrule
Ta--O(2) & x 2 & 1.877(4) & 1.871(4) \\
Ta--O(1) & x 2 & 1.938(4) & 1.945(3) \\
Ta--O(1) & x 2 & 2.301(4) & 2.297(4) \\
\midrule
Tb--O(2) & x 2 & 2.314(3) & 2.311(2) \\
Tb--O(1) & x 2 & 2.342(4) & 2.342(3) \\
Tb--O(2) & x 2 & 2.376(4) & 2.379(3) \\
Tb--O(1) & x 2 & 2.500(4) & 2.492(3) \\
\end{tabular}
\end{ruledtabular}
\end{table}

In addition, we were able to resolve and follow the evolution of the nuclear structure with temperature using PND data collected on the D1B beamline. No phase transitions were observed in the temperature range 3--300~K. The lattice parameters were constant between 3 and 50~K and then varied smoothly between 50 and 300~K. Small decreases in \textit{a} and $\beta$ on warming were offset by increases in $b$ and $c$, Fig.~S1. There were similar subtle changes to the atomic fractional coordinates between 50 and 300~K (Figs S2 and S3). 

\subsection{Bulk magnetic properties}
\subsubsection{Magnetic susceptibility}
\begin{figure*}[htbp]
\centering
\includegraphics[width=17.2cm]{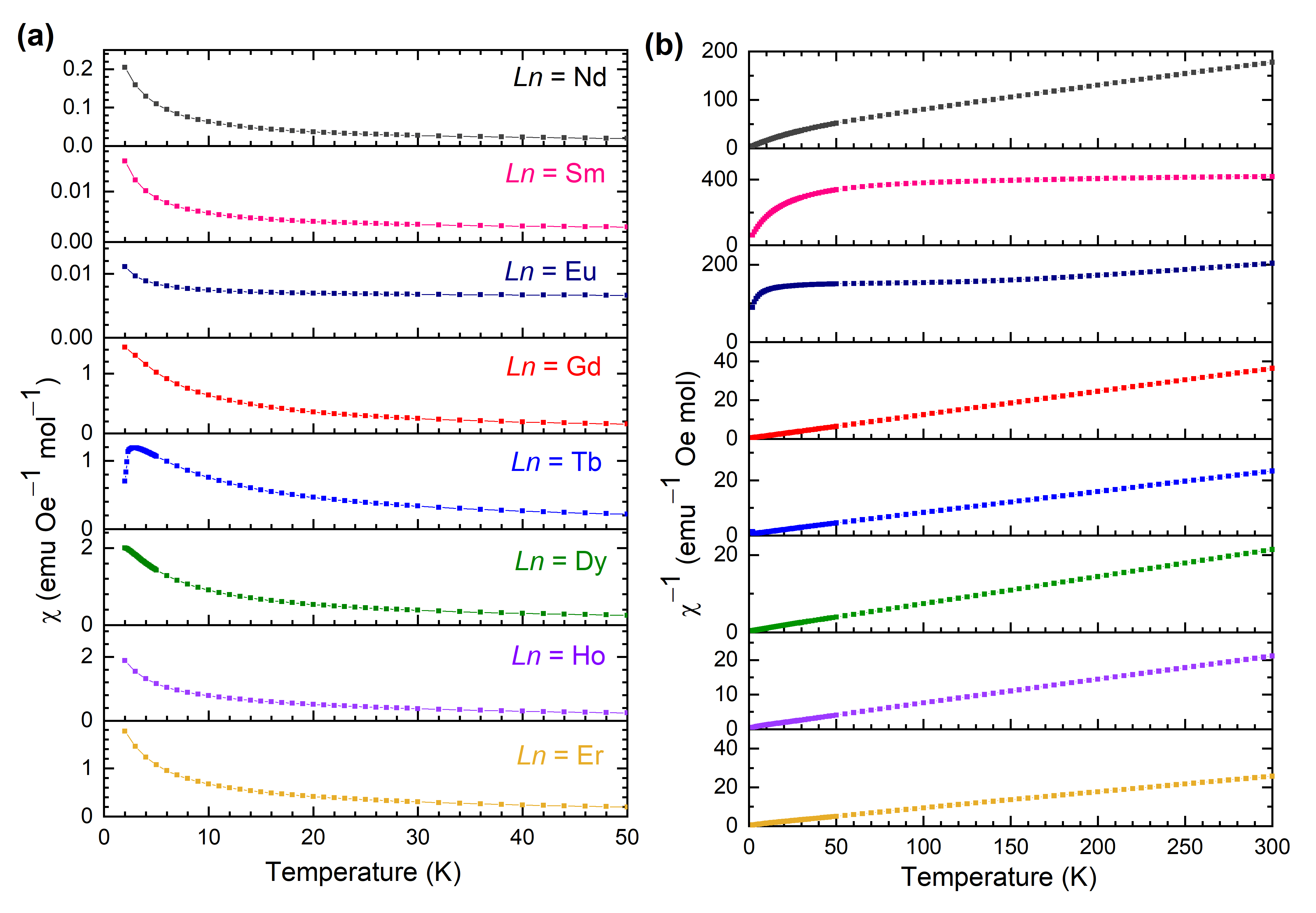}
\caption{(a) ZFC susceptibility $\chi$, (b) $(\chi-\chi_0)^{-1}$ as a function of temperature for the \lntao\ samples with \textit{Ln}~= Nd, Sm, Eu, Gd, Tb, Dy, Ho and Er.}
\label{fig:lntaomvtcombined}
\end{figure*}

The zero-field-cooled (ZFC) magnetic susceptibility at 500~Oe for each \lntao\ compound is plotted in Fig.~\ref{fig:lntaomvtcombined}(a). No sharp peaks in the susceptibility were observed for any of the eight compounds except \ch{TbTaO4}, which had a peak at 2.9~K. Field-cooled (FC) susceptibility at 500~Oe was also measured on \ch{TbTaO4} and found to be identical to the ZFC data, suggesting three-dimensional antiferromagnetic ordering without glassiness.

The susceptibility was fitted to the modified Curie-Weiss law:
\begin{equation}
\chi=\chi_0+\frac{C}{(T-\theta_\mathrm{CW})}
\label{eqn:modcw}
\end{equation}
where $\chi_0$ is the temperature-independent contribution to the susceptibility. Linear fitting of $(\chi-\chi_0)^{-1}$ against $T$, Fig.~\ref{fig:lntaomvtcombined}(b), was carried out in the temperature range 50--300~K. The effective magnetic moment was calculated from the experimental data using $\mu_\mathrm{eff}/\mu_\mathrm{B}=\sqrt{8C}$ and compared to the theoretical paramagnetic moment $g_J\sqrt{J(J+1)}$. The results are given in Table~\ref{table:lntaomagdatanew}. The magnitudes of the Curie-Weiss temperatures for \textit{Ln}~= Nd, Ho and Er are consistent with the results of Tsunekawa \textit{et al}. \cite{Tsunekawa1996}. The experimental magnetic moments are also in excellent agreement with the theoretical values, with the exceptions of \textit{Ln}~= Sm, Eu and Tb. The susceptibility of \ch{SmTaO4} at high temperatures shows a large contribution from temperature-independent paramagnetism. The calculated effective magnetic moment is 0.66~$\mu_\mathrm{B}$, slightly lower than the expected free ion value of 0.85~$\mu_\mathrm{B}$, likely owing to the large crystal field splitting in the $J=\nicefrac{5}{2}$ ground state multiplet of Sm$^{3+}$ \cite{Sanders2016,Sanders2017}. The negative Curie-Weiss temperature indicates antiferromagnetic interactions between adjacent Sm$^{3+}$ ions, as indeed is the case for all the remaining \lntao\ samples. The shape of the \ch{EuTaO4} susceptibility curve resembles that of other Eu$^{3+}$-containing ceramics and is believed to result from van Vleck paramagnetism, a second-order correction involving higher-lying $^7\mathrm{F}_1$--$^7\mathrm{F}_6$ states \cite{Nishimine2005,VijayaKumar2011,Sanders2016}. The inverse susceptibility plot is linear at 50--120~K and 200--300~K, but applying equation~\ref{eqn:modcw} produced unrealistically large values of the magnetic moment and Curie-Weiss temperature. Finally, the discrepancy between experimental and theoretical $\mu_\mathrm{eff}$ for \ch{TbTaO4} is $<3\sigma$ but larger than the discrepancy for \textit{Ln}~= Nd, Gd, Dy, Ho and Er, likely because of magnetic correlations developing in the lower temperature range, since \ch{TbTaO4} is the only compound to order above 2~K.

\begin{table*}[htbp]
\caption{Bulk magnetic properties of $M$-\lntao, \textit{Ln}~= Nd, Sm, Eu, Gd, Tb, Dy, Ho, Er.}
\label{table:lntaomagdatanew}
\begin{ruledtabular}
\begin{tabular}{c c c c c c c c}
\textit{Ln} & $g_J\sqrt{J(J+1)}$ & $\chi_0$ (emu mol$^{-1}$ Oe$^{-1}$) & $C$ (emu mol$^{-1}$ Oe$^{-1}$ K) & $\theta_\mathrm{CW}$ (K) & $\mu_\mathrm{eff}$ ($\mu_\mathrm{B}$) &  $g_J.J$ & $M_\mathrm{2K,9T}$ ($\mu_\mathrm{B}$/f.u.) \\
\midrule
Nd & 3.62 & $8\times10^{-4}$ & 1.65(3) & --40.5(8) & 3.63(7) & 3.29 & 1.312(26) \\
Sm & 0.85 & $2.2\times10^{-3}$ & 0.0548(11) & --29.8(6) & 0.662(13) & 0.71 & 0.0749(15) \\
Eu & 0 & N/A & N/A & N/A & N/A & 0 & 0.1037(21) \\ 
Gd & 7.94 & $6\times10^{-4}$ & 8.14(16) & --2.77(6) & 8.07(16) & 7 & 6.88(14) \\
Tb & 9.72 & $8\times10^{-4}$ & 12.96(26) & --9.49(19) & 10.18(20) & 9 & 5.48(11) \\
Dy & 10.65 & 0 & 14.33(29) & --6.88(14) & 10.71(21) & 10 & 6.00(12) \\ 
Ho & 10.61 & $1.9\times10^{-3}$ & 13.97(28) & --7.84(16) & 10.57(21) & 10 & 6.72(13) \\
Er & 9.58 & $2.6\times10^{-3}$ & 11.15(22) & --7.43(15) & 9.44(19) & 9 & 5.98(12) \\
\end{tabular}
\end{ruledtabular}
\end{table*}

\subsubsection{Isothermal magnetization}
\begin{figure}[htbp]
\centering
\includegraphics[width=8.6cm]{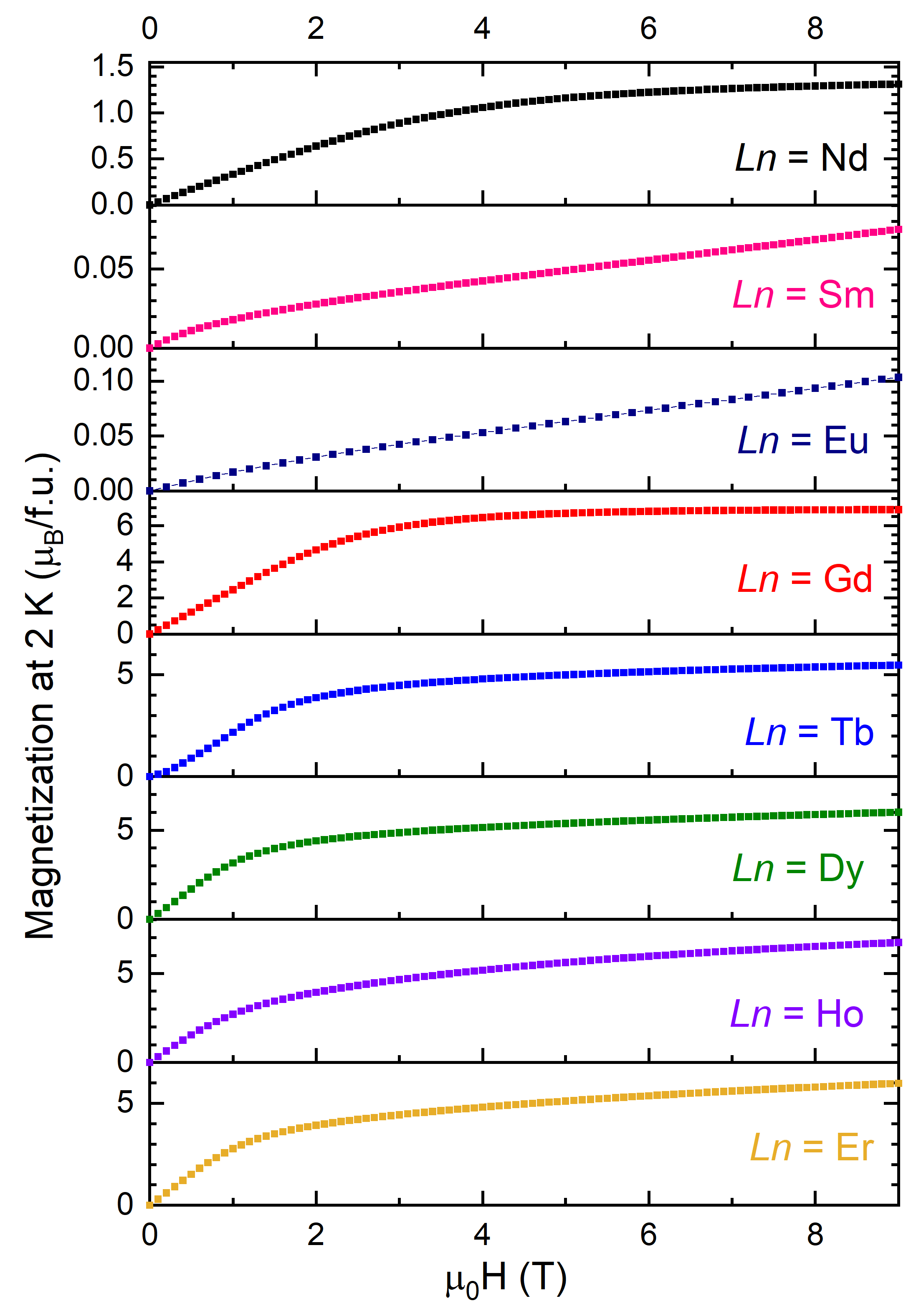}
\caption{Magnetization as a function of applied field for the \lntao\ samples with \textit{Ln}~= Nd, Sm, Eu, Gd, Tb, Dy, Ho and Er.}
\label{fig:lntaomvhall}
\end{figure}

Fig.~\ref{fig:lntaomvhall} shows the isothermal magnetization at 2~K for the \lntao\ compounds. For the samarium and europium compounds, the magnetization plots initially curve upwards then become linear above 3~T without saturating. In all other samples the magnetization, plotted in units of Bohr magnetons per formula unit ($\mu_\mathrm{B}$/f.u.), tends towards a saturation value $M_\mathrm{sat}$ at high field. The expected value of $M_\mathrm{sat}$ depends on both the identity of the lanthanide ion and the extent of single-ion anisotropy -- the tendency for a spin to align along a particular local axis or local plane. For example, compounds containing Gd$^{3+}$ typically display Heisenberg-type behavior with saturation at the maximum value of $g_J.J=7$~$\mu_\mathrm{B}$/f.u., while systems with Ising (easy-axis) or $XY$ (easy-plane) behavior are expected to saturate around $g_J.J/2$ or $2g_J.J/3$ respectively. However, individual systems may vary from these values depending on the local symmetry of the lanthanide ion coordination environment \cite{Bramwell2000,Dixey2018}. The experimental values of $M_\mathrm{2K,9T}$ for each compound and the calculated $g_J.J$ for each lanthanide ion are given in Table~\ref{table:lntaomagdatanew}.

Bulk powder averaging limits the information that can be obtained on crystal field effects from the $M(H)$ data. However, the $M_\mathrm{sat}$ values do indicate that the compounds with \textit{Ln}~= Nd, Tb, Dy, Ho and Er all show some degree of local anisotropy. Further experiments such as neutron diffraction would be needed to investigate this further, although neutron absorption would make it difficult to measure the anisotropy in \ch{DyTaO4} without an isotopically enriched sample. A previous study on large single crystals of \lntao\ (\textit{Ln}~= Nd, Ho, Er) also found substantial anisotropy in the magnetic susceptibility measured along the different crystal axes \cite{Tsunekawa1996}. In that study, the plots of inverse susceptibility along each crystal axis have the same gradient but different $x$-intercepts, i.e.~Curie-Weiss temperatures: for example, \ch{NdTaO4}, which showed the greatest anisotropy, had $\theta_\mathrm{CW}=-7$, --56 and --52~K along the $a$, $b$ and $c$ axes respectively. This illustrates the importance of single-crystal studies for gaining a better understanding of the magnetostructural anisotropy in compounds such as \lntao.

\subsubsection{Specific heat}
\begin{figure}[htbp] 
\centering
\includegraphics[width=8.6cm]{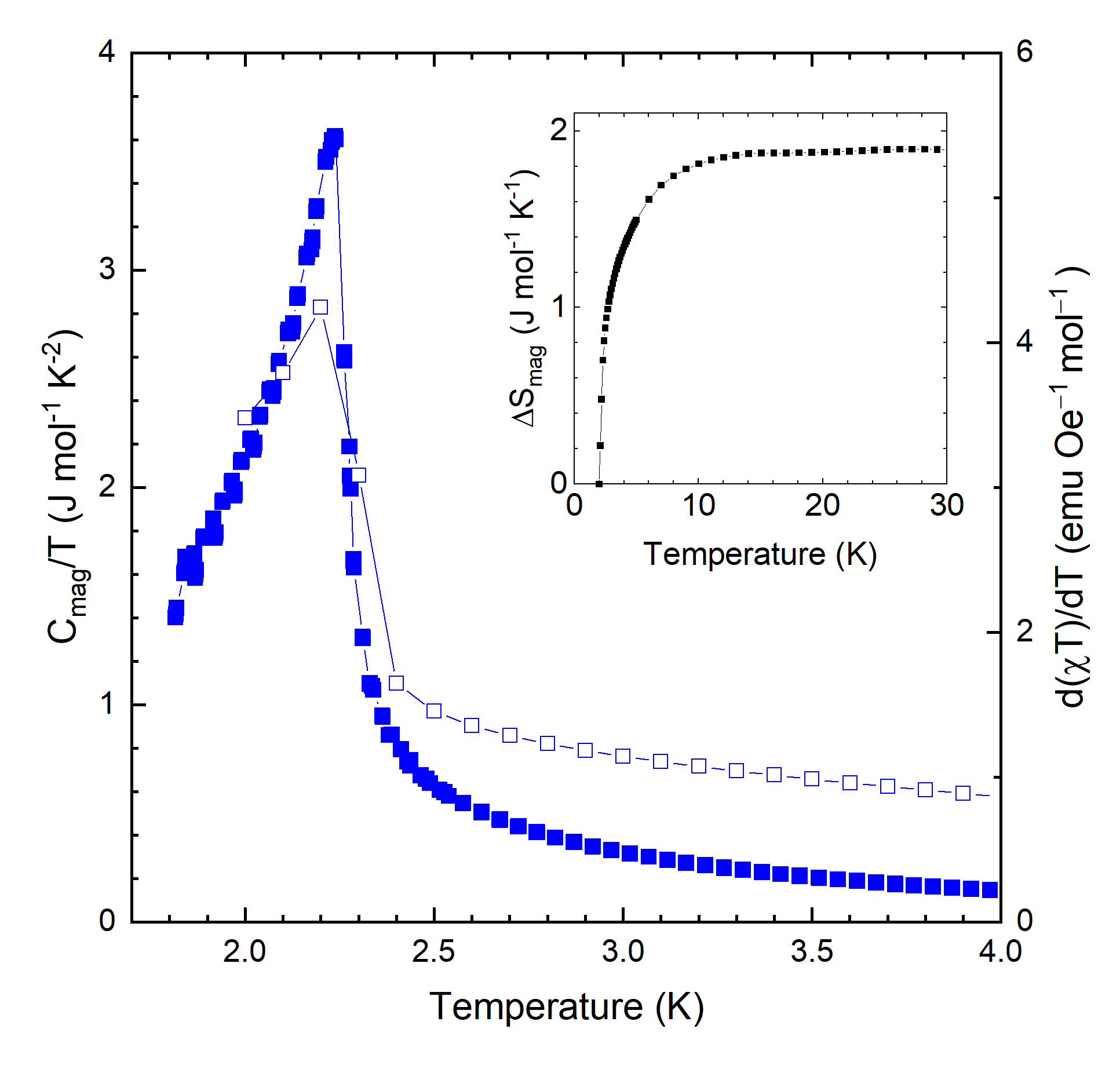}
\caption{Magnetic heat capacity (filled symbols) and $d(\chi T)/dT$ vs $T$ (open symbols) for \ch{TbTaO4}. Inset: magnetic entropy $\Delta S_\mathrm{mag}$ obtained by integration of $C_\mathrm{mag}/T$.}
\label{fig:tbmaghcandchit}
\end{figure}

The magnetic heat capacity for \ch{TbTaO4}was obtained from the total heat capacity by subtraction of the estimated lattice contribution using Debye fitting ($\theta_\mathrm{D}=370$~K). The subtraction is shown in the Supplemental Material. The magnetic heat capacity shows a sharp $\lambda$-type transition at $T=2.25$~K, where there is a corresponding feature in the plot of $d(\chi T)/dT$ \cite{Fisher1962}, Fig.~\ref{fig:tbmaghcandchit}. This provides further evidence for three-dimensional antiferromagnetic ordering as deduced from the susceptibility data. The magnetic entropy associated with the transition was obtained by integration of the heat capacity curve over the full temperature range (1.8--30~K) and found to approach 2~J~mol$^{-1}$~K$^{-1}$ (Fig.~\ref{fig:tbmaghcandchit}, inset). Since the expected maximum entropy is $R\ln2=5.76$~J~mol$^{-1}$~K$^{-1}$ for Ising spins with effective spin of $\nicefrac{1}{2}$, the remaining entropy change is assumed to occur below the lowest temperature measured (1.8~K) which has non-zero $C_\mathrm{mag}/T$ and is close to $T_\mathrm{N}$.

\subsection{Magnetic structure of \ch{TbTaO4}}
\begin{figure}[htbp]
\centering
\includegraphics[width=8.6cm]{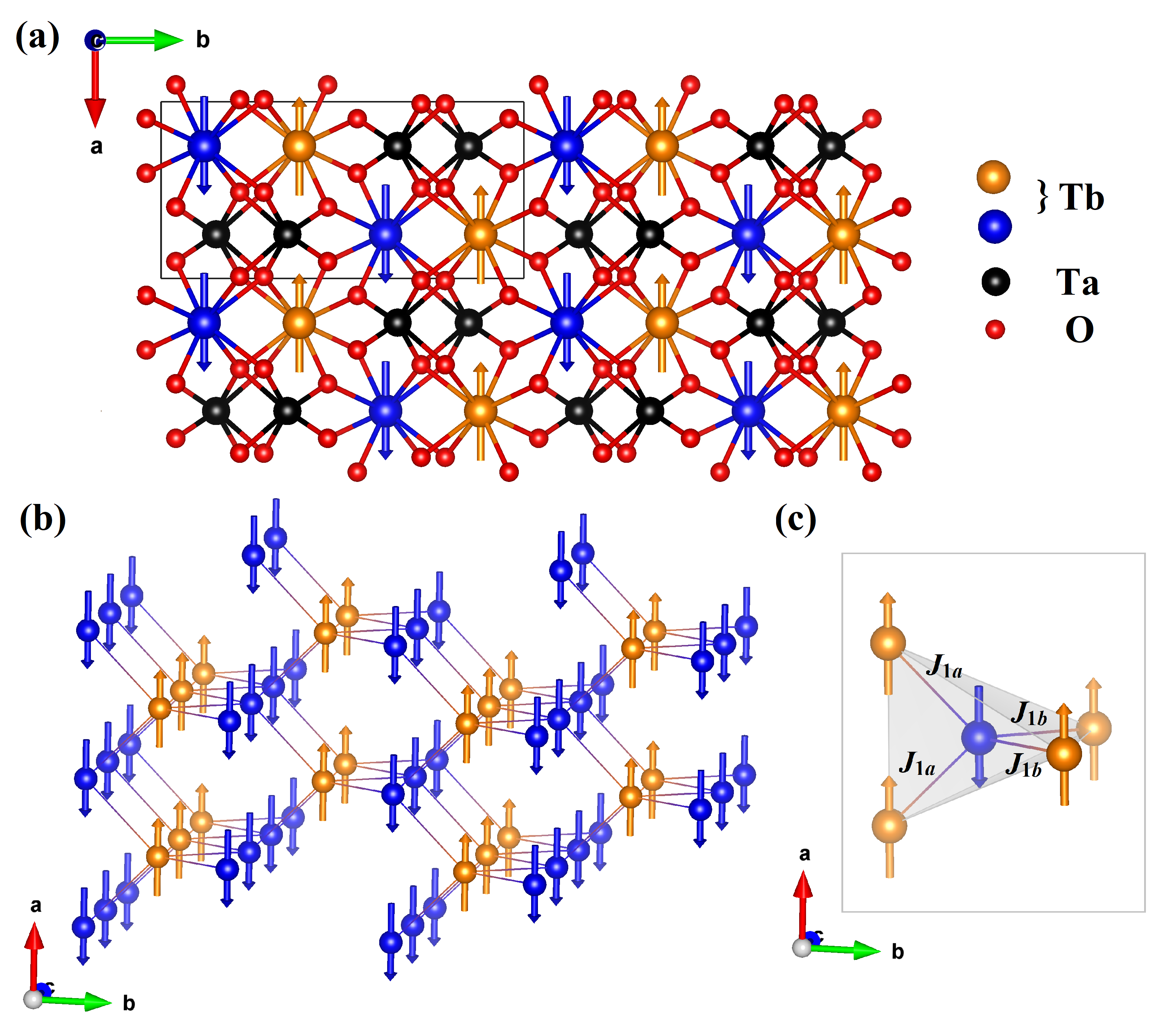}
\caption{Magnetic structure of \ch{TbTaO4} from refinement at 1.5~K: (a) all atoms, (b) Tb spins only, (c) nearest-neighbor interactions.}
\label{fig:tbtaomagstruct}
\end{figure}

We obtained high-resolution PND data for the magnetic phase of \ch{TbTaO4} at 1.5~K using the D2B beamline at the ILL \cite{Kelly2021ILL}. Variable-temperature PND was also carried out on the D1B beamline in order to track the evolution of the magnetic structure with temperature. On cooling below $T=2.3$~K, magnetic Bragg peaks were observed to appear and increase in intensity as the temperature was lowered. No discernible diffuse scattering was observed above this temperature. The magnetic peaks could be indexed to a commensurate magnetic cell with $\vec{k}=0$ in the magnetic space group 14.77 ($P2_1$$'/c$). Refinement of the magnetic structure using \textsc{Topas} (Fig.~\ref{fig:tbtaopnd}(b)) shows Tb moments parallel to the \textit{a}-axis in $A$-type antiferromagnetic order: the moments coalign within the $ac$ plane, forming ferromagnetic slabs coupled antiferromagnetically along $b$, Fig.~\ref{fig:tbtaomagstruct}(a). The structure is similar to that of \ch{NaCeO2} which has the same $A$-type order but Ce$^{3+}$ moments aligned along the tetragonal \textit{c}-axis \cite{Bordelon2021}. Further details of the magnetic structure may be found in the Supplemental Material.

\begin{figure}[htbp]
\centering
\includegraphics[width=8.6cm]{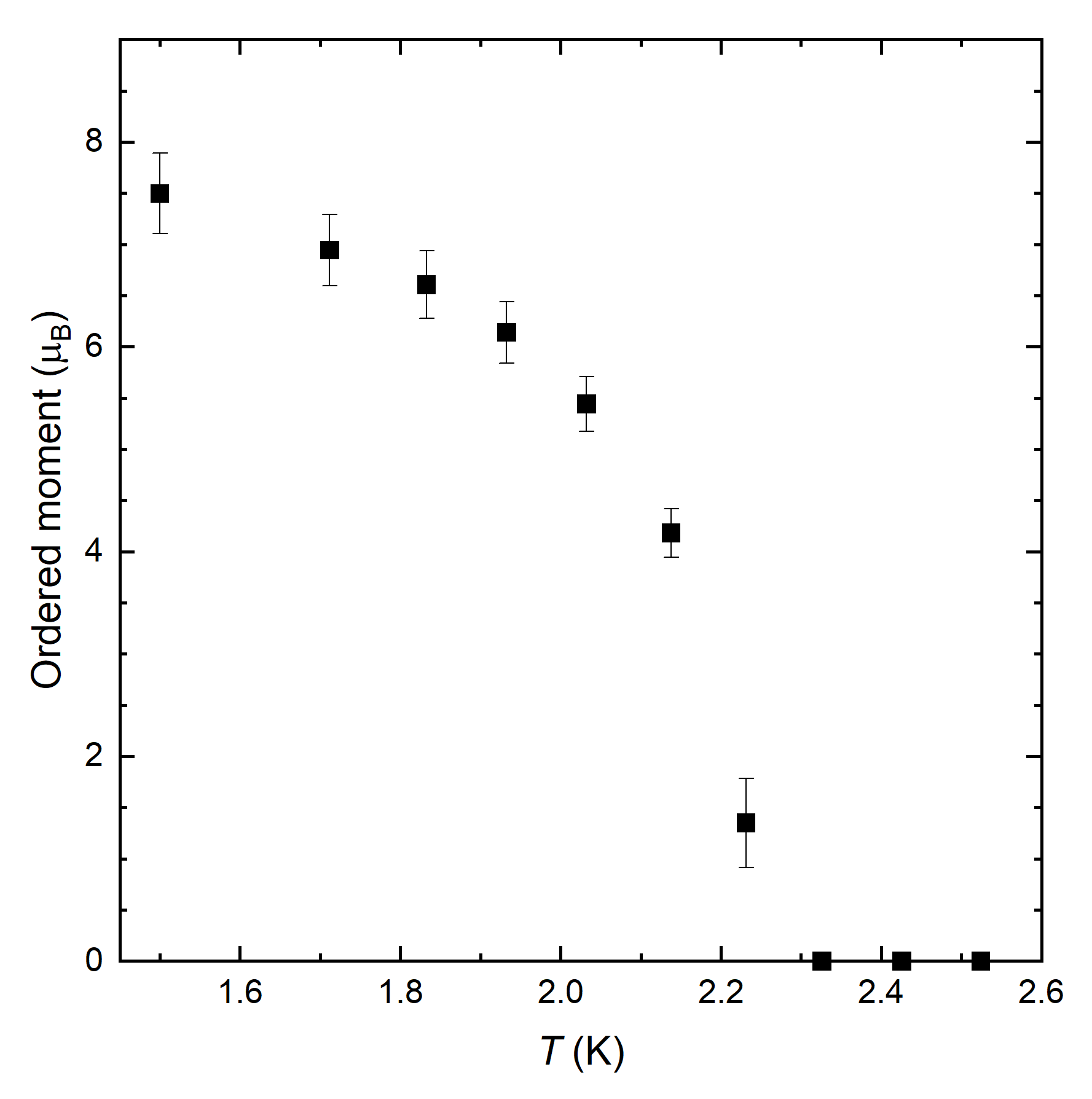}
\caption{Evolution of the Tb$^{3+}$ ordered magnetic moment as a function of temperature from variable-temperature PND.}
\label{fig:tbmomenttemp}
\end{figure}

The refined ordered moment is plotted as a function of temperature in Fig.~\ref{fig:tbmomenttemp}, showing good agreement with the transition temperature found by heat capacity and magnetic susceptibility. At 1.5~K the moment of 7.5(4)~$\mu_\mathrm{B}$/Tb$^{3+}$ is slightly below the expected value of 9~$\mu_\mathrm{B}$.

The nearest-neighbor superexchange in \ch{TbTaO4} follows Tb--O--Tb pathways. These pathways may be divided into $J_{1a}$ and $J_{1b}$ according to the different Tb--O bond lengths. Figs~\ref{fig:tbtaomagstruct}(b) and \ref{fig:tbtaomagstruct}(c) highlight these two interactions: $J_{1a}$ vectors (shorter) in the $ab$ plane and $J_{1b}$ vectors (longer) in the $bc$ plane. In the mean-field approximation, the average exchange interaction $\bar{J_1}$ may be calculated as $\bar{J_1}=[3k_\mathrm{B}\theta_\mathrm{CW}]/[2nJ(J+1)]$, where $J$ is the spin quantum number and $n$ the number of nearest-neighbor spins \cite{Ramirez1994}. Using an effective spin of $\nicefrac{1}{2}$ for the Tb$^{3+}$ ion \cite{Mukherjee2017a} we estimate $\bar{J_1}\approx4.7$~K, of the same order as the N\'{e}el temperature. However, the mean-field approximation may not be completely valid given the significant single-ion anisotropy observed in $M(H)$ data. Growth of large single crystals of all \lntao\ would be important for further investigations and effective modeling of the crystal electric field and magnetic anisotropy.

\section{Discussion}
We report the magnetic behavior of a family of \textit{Ln}$^{3+}$ oxides containing a stretched diamond lattice of magnetic ions. Other such materials include the alkali metal-lanthanide oxides \ch{Na\textit{Ln}O2} (\textit{Ln}~= Ce, Nd, Sm, Eu, Gd) and \ch{Li\textit{Ln}O2} (\textit{Ln}~= Er, Yb), which crystallize in the tetragonal space group $I4_1/amd$ \cite{Hashimoto2003,Bordelon2021,Bordelon2021a}. The observation of the stretched diamond lattice in monoclinic \lntao\ materials provides a new opportunity to study the interplay of the crystal electric field with competing $J_1$ and $J_2$ magnetic interactions. In particular, while \ch{TbTaO4} exhibits long-range $A$-type antiferromagnetic order similar to that of \ch{NaCeO2} \cite{Bordelon2021}, the other materials with \textit{Ln}~= Nd--Er do not order at $T\geq2$~K. The absence of ordering above 2~K, in contrast to e.g.~\ch{Na\textit{Ln}O2} (Ce, Gd antiferromagnetic; Nd ferromagnetic \cite{Hashimoto2003}) suggests the potential for novel magnetic states at low temperature.

The Curie-Weiss temperatures for \lntao\ listed in table~\ref{table:lntaomagdatanew} are of the order of 2--10~K for \textit{Ln}~= Gd--Er, and much larger (41~K) for \ch{NdTaO4}, so we expect to observe magnetic correlations in this temperature range. The data for \textit{Ln}~= Nd, Gd, Tb, Dy, Ho and Er were plotted according to a dimensionless form of the Curie-Weiss equation,
\begin{equation}
\frac{C}{(\chi-\chi_0)|\theta|}=\frac{T}{|\theta|}+1
\label{eqn:melotcw}
\end{equation}
in order to look for signs of being near to a magnetic ordering transition, since deviation from linearity at low temperatures is an indication of short-range magnetic correlations \cite{Melot2009,Koskelo2022}. The plots are shown in the Supplemental Material (Fig.~S5) and indicate that all materials are developing weak correlations only below $T/|\theta|\approx1$ as expected. However, even \ch{NdTaO4} with the largest $|\theta|$ fails to order above 2~K, which illustrates how frustration suppresses magnetic ordering to lower temperatures in \lntao.

\begin{figure*}[htbp]
\centering
\includegraphics[width=17.2cm]{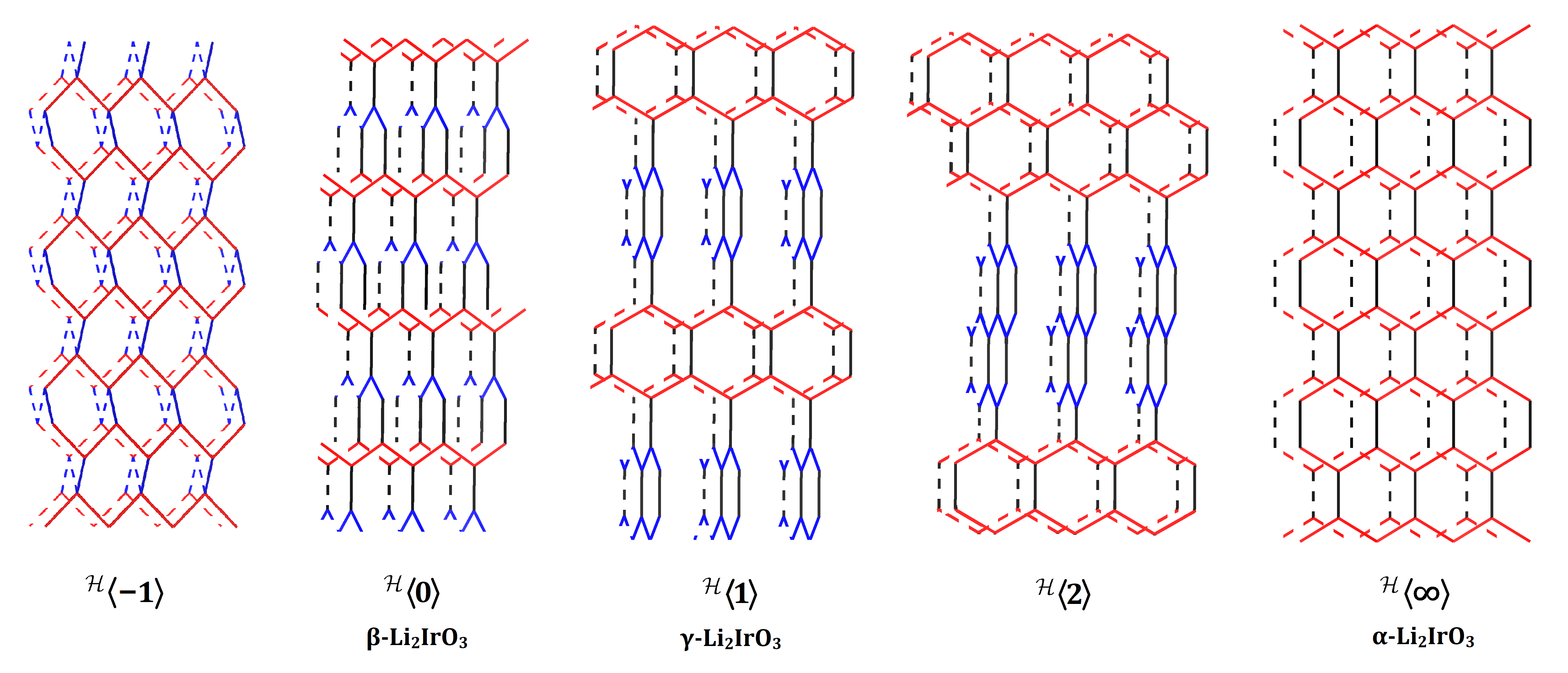}
\caption{The $N=-1$, 0, 1, 2, and $\infty$ members of the harmonic honeycomb series, where $N$ represents the number of complete honeycomb rows along the $c$-axis before a change of orientation of the honeycomb planes. After Modic~\textit{et al}. \cite{Modic2014}.}
\label{fig:honeycombs}
\end{figure*}

The diamond lattice can also be viewed as a truncated version of the $^{\mathcal{H}}\langle0\rangle$ `hyperhoneycomb' structure of $\beta$-\ch{Li2IrO3}, using nomenclature for the so-called harmonic honeycomb series \cite{Modic2014}. This `truncation' is carried out by removing the black links (parallel to $c$) from the $N=0$ structure, as shown in Fig.~\ref{fig:honeycombs}. $N$ stands for the number of complete hexagonal rows along the $c$-axis before a change of orientation of the hexagons. Alternatively, $N+1$ is the number of black ($c$-axis) links between changes of orientation \cite{Kimchi2014}; as such, we propose the notation $^{\mathcal{H}}\langle-1\rangle$ for the diamond lattice. Magnetic frustration in materials with the $^{\mathcal{H}}\langle-1\rangle$ connectivity is probable in crystal symmetries lower than cubic, and has previously been investigated in \ch{NaCeO2} \cite{Bordelon2021} and \ch{LiYbO2} \cite{Bordelon2021a}.

Like the materials in the wider series $^{\mathcal{H}}\langle N\rangle$, which include \ch{YbCl3} \cite{Sala2019} and the $\alpha$, $\beta$ and $\gamma$ polymorphs of \ch{Li2IrO3} \cite{Kimchi2014}, magnetism on the $^{\mathcal{H}}\langle-1\rangle$ lattice is governed by the interplay of nearest-neighbor and next-nearest-neighbor interactions. Describing the stretched diamond lattice as $^{\mathcal{H}}\langle-1\rangle$ provides a useful framework to draw parallels between different materials within the harmonic honeycomb series.

There are two parameters which are useful for comparing the level of stretching in different stretched diamond or $^{\mathcal{H}}\langle-1\rangle$ lattices. The first is the angle or angles around each lattice vertex. In an ideal cubic diamond lattice, all these angles are equal at 109.5\degree. When the diamond lattice is distorted, the number of different angles increases: there are two in tetragonal, three in hexagonal, and four in monoclinic symmetry. We compare the average deviation from ideal tetrahedral geometry by defining a parameter $d_\mathrm{a}$, the angle distortion index, as follows:
\begin{equation}
d_\mathrm{a}=\frac{\phi_\mathrm{max}-\phi_\mathrm{min}}{\bar{\phi}}
\label{eqn:angleparam}
\end{equation}
where $\phi_\mathrm{max}$ and $\phi_\mathrm{min}$ are the largest and smallest angles respectively, and $\bar{\phi}$ is the mean angle. Secondly, we define a bond distortion index $d_\mathrm{b}$ in a similar way:
\begin{equation}
d_\mathrm{b}=\frac{r_2-r_1}{\frac{1}{2}(r_1+r_2)}
\label{eqn:distortion}
\end{equation}
where $r_1$ and $r_2$ are the two `nearest-neighbor' interaction distances (red and blue in Fig.~\ref{fig:honeycombs}). These distances are equal in the case of the undistorted (cubic) or the tetragonal or hexagonal stretched diamond lattices, but not in monoclinic materials such as the tantalates. Table~\ref{table:distortion} lists the distortion indices for several materials containing this magnetic lattice. Values were calculated using the program \textsc{Vesta} \cite{Momma2011} to examine the published crystal structures.

\begin{table*}[htbp]
\caption{Distortion indices and magnetic behavior of materials with the stretched diamond lattice.}
\label{table:distortion}
\resizebox{\textwidth}{!}{
\begin{ruledtabular}
\begin{tabular}{c c c c c c c}
Crystal symmetry & Formula & Space group & $d_\mathrm{a}$ (\%) & $d_\mathrm{b}$ (\%) & $T_\mathrm{N}$ (K) & Magnetic structure \\ 
\midrule
Cubic & \ch{CoRh2O4} \cite{Ge2017} & $Fd\bar{3}m$ & 0 & 0 & 25 & N\'{e}el AFM $\parallel\langle100\rangle$ \\ 
& \ch{MnAl2O4} \cite{Tristan2005} & $Fd\bar{3}m$ & 0 & 0 & 40 & Canted AFM \\ 
& \ch{FeAl2O4} \cite{Tristan2005} & $Fd\bar{3}m$ & 0 & 0 & 12 & Spin glass \\
& \ch{CoAl2O4} \cite{Tristan2005} & $Fd\bar{3}m$ & 0 & 0 & 4.8 & Spin glass \\
& \ch{CuAl2O4} \cite{Cho2020} & $Fd\bar{3}m$ & 0 & 0 & $<0.4$ & No long range order \\
& \ch{FeSc2S4} \cite{Fritsch2004} & $Fd\bar{3}m$ & 0 & 0 & $<0.05$ & Spin-orbital liquid \\
& \ch{MnSc2S4} \cite{Bergman2007} & $Fd\bar{3}m$ & 0 & 0 & 2.3, 1.9 & Long-range spiral order \\
\midrule
Tetragonal & \ch{CuRh2O4} \cite{Ge2017} & $I4_1/amd$ & 7 & 0 & 24 & Incommensurate helical order \\
& \ch{NiRh2O4} \cite{Chamorro2018} & $I4_1/amd$ & 3 & 0 & $<0.1$ & No long range order  \\ 
& \ch{NaCeO2} \cite{Bordelon2021} & $I4_1/amd$ & 42 & 0 & 3.18 & N\'{e}el AFM $\parallel c$ \\ 
& \ch{NaNdO2} \cite{Hashimoto2003} & $I4_1/amd$ & 41 & 0 & 2.4 & FM \\ 
& \ch{NaGdO2} \cite{Hashimoto2003} & $I4_1/amd$ & 40 & 0 & 2.4 & AFM \\
& \ch{LiYbO2} \cite{Bordelon2021a} & $I4_1/amd$ & 41 & 0 & 0.45 & Incommensurate helical order \\ 
& \ch{KRuO4} \cite{Marjerrison2016} & $I4_1/a$ & 42 & 0 & 22.4 & N\'{e}el AFM $\parallel c$ \\ 
& \ch{KOsO4} \cite{Injac2019} & $I4_1/a$ & 39 & 0 & 35 & N\'{e}el AFM $\parallel c$ \\ 
\midrule
Hexagonal & $\beta$-\ch{KTi(C2O4)2}$\cdot$2\ch{H2O} \cite{Abdeldaim2020} & $P6_222$ & 34 & 0 & 28 & Coplanar AFM \\
\midrule
Monoclinic & \ch{NdTaO4} & $I2/a$ & 42 & 2.9 & $<2$ & No long range order \\ 
& \ch{GdTaO4} & $I2/a$ & 41 & 2.6 & $<2$ & No long range order \\ 
& \ch{TbTaO4} & $I2/a$ & 41 & 2.2 & 2.25 & N\'{e}el AFM $\parallel a$ \\  
& \ch{DyTaO4} & $I2/a$ & 41 & 4.1 & $<2$ & No long range order \\ 
& \ch{HoTaO4} & $I2/a$ & 41 & 4.1 & $<2$ & No long range order \\ 
& \ch{ErTaO4} & $I2/a$ & 41 & 3.9 & $<2$ & No long range order \\ 
\cline{2-7}
& \ch{NdNbO4} \cite{Cashion1968,Tsunekawa1993a} & $I2/a$ & 40 & 1.1 & $<1$ & No long range order \\ 
& \ch{GdNbO4} \cite{Cashion1968,Trunov1982} & $I2/a$ & 39 & 0.7 & 1.67 & AFM \\
& \ch{TbNbO4} \cite{Cashion1968,Keller1962} & $I2/a$ & 39 & 1.0 & 1.82 & AFM \\ 
& \ch{DyNbO4} \cite{Cashion1968,Keller1962} & $I2/a$ & 39 & 0.3 & 1.6 & AFM \\
& \ch{HoNbO4} \cite{Cashion1968,Tsunekawa1993a} & $I2/a$ & 39 & 1.1 & $<1$ & No long range order \\
& \ch{ErNbO4} \cite{Cashion1968,Keller1962} & $I2/a$ & 38 & 0.6 & $<1.3$ & No long range order \\ 
& \ch{YbNbO4} \cite{Cashion1968,Tsunekawa1993a} & $I2/a$ & 38 & 1.2 & $<1.3$ & No long range order \\
\cline{2-7}
& \ch{Pr(BO2)3} \cite{Mukherjee2017} & $C2/c$ & 61 & 12.1 & -- & Singlet ground state \\
& \ch{Nd(BO2)3} \cite{Mukherjee2017} & $C2/c$ & 61 & 12.0 & $<0.4$ & No long range order \\ 
& \ch{Gd(BO2)3} \cite{Mukherjee2017} & $C2/c$ & 64 & 12.6 & 1.1 & AFM \\ 
& \ch{Tb(BO2)3} \cite{Mukherjee2017} & $C2/c$ & 66 & 13.1 & 1.95, 1.05 & Undetermined \\
\end{tabular}
\end{ruledtabular}}
\end{table*}

Table~\ref{table:distortion} highlights the fact that the magnetic ordering of \ch{TbTaO4} occurs at a similar temperature to \ch{NaNdO2} and \ch{NaGdO2}, despite the higher symmetry of \ch{Na\textit{Ln}O2}, whereas the remaining \lntao\ do not order above 2~K. The effect of distortion on magnetism on this lattice is clearly important but the precise mechanism is currently unclear. The suppression of ordering temperatures in certain materials likely depends on a combination of the distortion and the identity of the magnetic ion, i.e.~anisotropy and crystal field splitting, both of which would influence the superexchange and/or dipolar interactions. Future experiments including inelastic neutron scattering or polarized neutron diffraction, especially on single-crystal samples, would be valuable for determining the extent to which distortion affects the crystal electric field and hence the magnetic properties of these materials.

Finally, we note that \ch{LaTaO4}, while not forming the fergusonite structure type, nonetheless displays rich structural phase behavior including an incommensurate--commensurate phase transition at 483~K coinciding with a dielectric anomaly \cite{Howieson2020,Howieson2021}. Dias \textit{et al}.~have made comparative dielectric measurements on \ch{LaTaO4} ($P2_1/c$), \ch{NdTaO4} ($I2/a$), and \ch{DyTaO4} and \ch{LuTaO4} ($P2/a$). Despite the differences in structure, the dielectric constants of the Nd, Dy and Lu tantalates are $\leq25$~\%\ smaller than that of \ch{LaTaO4} and still within the range of useful microwave materials \cite{Dias2017}. Considering that Nd$^{3+}$ and Dy$^{3+}$ have non-zero magnetic moments, there is a possibility of coupling between magnetic and electric properties which should be investigated in these and the other magnetic tantalates.

\section{Conclusions}
Polycrystalline samples of \lntao\ (\textit{Ln}~= Nd, Sm--Er, Y) in the monoclinic $M$, or fergusonite, structure type have been synthesized using a ceramic procedure. The trivalent lanthanide ions in the crystal structure form a three-dimensional net equivalent to an elongated or `stretched' diamond lattice. This lattice can also be considered a truncated form of the hyperhoneycomb lattice $^{\mathcal{H}}\langle 0\rangle$, part of the harmonic honeycomb series, and is therefore denoted $^{\mathcal{H}}\langle-1\rangle$. Bulk magnetic characterization of the tantalate samples confirms a previous literature result for \textit{Ln}~= Nd, Ho, Er, and reveals that the remaining compounds do not order above 2~K with the exception of \ch{TbTaO4}, which has $T_\mathrm{N}=2.25$~K. High-resolution PND was used to examine the paramagnetic and magnetic phases of \ch{TbTaO4}, revealing that it forms a commensurate $\vec{k}=0$ magnetic unit cell. The Tb moments lie parallel to the \textit{a}-axis in $A$-type antiferromagnetic order. Future work will include specific heat measurements at $T<2$~K in order to search for further magnetic transitions and investigate the magnetic ground states of the remaining tantalates.

\begin{acknowledgments}
N.D.K.~thanks Sundeep Vema for carrying out the reactions at 1600~\degree C, Farheen Sayed for the SPS experiments and James Analytis for technical advice on producing Fig.~\ref{fig:honeycombs}. We acknowledge funding from the EPSRC (EP/T028580/1, EP/R513180/1, EP/M000524/1).

Supplementary tables and figures are available at (DOI). Neutron diffraction data are available at reference \cite{Kelly2021ILL}. The authors gratefully acknowledge the technical and human support provided at the Institut Laue-Langevin (ILL), Grenoble. Additional data related to this publication are available in the Cambridge University Repository at (DOI).
\end{acknowledgments}


%

\end{document}